\documentclass[english,aps,aip,jcp,english,dvipsnames,preprint]{revtex4-1}

\usepackage[T1]{fontenc}
\usepackage[latin9]{inputenc}
\usepackage{color}
\usepackage{babel}
\usepackage{array}
\usepackage{amsmath}
\usepackage{amssymb}
\usepackage{graphicx}
\usepackage{esint}
\usepackage[unicode=true,pdfusetitle,
 bookmarks=true,bookmarksnumbered=false,bookmarksopen=false,
 breaklinks=false,pdfborder={0 0 1},backref=false,colorlinks=true]
 {hyperref}
\usepackage{breakurl}

\makeatletter

\newcommand{\noun}[1]{\textsc{#1}}
\providecommand{\tabularnewline}{\\}

\def\part#1{\left(#1\right)}

\usepackage{amsthm}
\usepackage{graphics}
\usepackage{times}
\usepackage{bm}
\usepackage{hyperref}
\usepackage{color}

\RequirePackage[dvipsnames]{xcolor}
\definecolor{RoyalBlue}{cmyk}{1, 0.80, 0, 0}

\hypersetup{
colorlinks=true,
breaklinks=true,
hyperfootnotes=true,
urlcolor=RoyalBlue,
citecolor=RoyalBlue, 
linkcolor=RoyalBlue, 
}

\makeatother

\begin{document}

\title{Mixed Semiclassical Initial Value Representation Time-Averaging Propagator
for Spectroscopic Calculations}

\author{Max \surname{Buchholz}}

\affiliation{Institut für Theoretische Physik, Technische Universität Dresden,
01062 Dresden, Germany and Max-Planck-Institut f\"ur Physik Komplexer
Systeme, N\"othnitzer Str. 38, 01187 Dresden, Germany}

\author{Frank \surname{Grossmann}}

\affiliation{Institut für Theoretische Physik, Technische Universität Dresden,
01062 Dresden, Germany }

\author{Michele \surname{Ceotto}}

\affiliation{Dipartimento di Chimica, Università degli Studi di Milano, via C.
Golgi 19, 20133 Milano, Italy}

\email{michele.ceotto@unimi.it}

\begin{abstract}
A mixed semiclassical initial value representation expression for
spectroscopic calculations is derived. The formulation takes advantage
of the time-averaging filtering and the hierarchical properties of
different trajectory based propagation methods. A separable approximation
is then introduced that greatly reduces (about an order of magnitude)
the computational cost compared with a full Herman-Kluk time-averaging
semiclassical calculation for the same systems. The expression is
exact for the harmonic case and it is tested numerically for a Morse
potential coupled to one or two additional harmonic degrees of freedom.
Results are compared to full Herman-Kluk time-averaging calculations
and exact quantum wavepacket propagations. We found the peak positions
of the mixed semiclassical approximations to be always in very good
agreement with full quantum calculations, while overtone peak intensities
are lower with respect to the exact ones. Given the reduced computational
effort required by this new mixed semiclassical approximation, we
believe the present method to make spectroscopic calculations available
for higher dimensional systems than accessible before.
\end{abstract}
\maketitle

\section{Introduction\label{sec:Introduction}}

Molecular spectra, and spectroscopic signals in general, contain all
quantum mechanical information connected to molecular motion, even
for increasingly complex systems. Most frequently, however, the full
amount of information is not useful and not necessary. Actually, it
would be convenient to be able to select with precision a certain
amount of spectroscopic information that is related to a subset of
degrees of freedom, which one is most interested in. In other words,
since not all degrees of freedom are equally important, many relevant
molecular properties of chemical systems can be rationalized in terms
of few main degrees of freedom, usually called the ``system''. These
are coupled to an environment of many other degrees of freedom, the
``bath'', which is not directly responsible for the physical properties
in question.

In a time-dependent approach to spectroscopy, exact quantum mechanical
methods to determine the dynamics of the complete dynamical system
are inaccessible for most real-life applications. A common strategy
therefore is to employ an accurate quantum propagator for the time-evolution
of the system degrees of freedom and a lower accuracy propagation
scheme for the bath ones. This should be done without enforcing any
artificial and arbitrary decoupling between the system and the bath.
Semiclassical Initial Values Representation (SC-IVR) molecular dynamics
\cite{Miller_avd_74,Miller_PNAS,Kay_review,Heller_IVR} is a valuable
tool for exploiting this strategy, since it offers a hierarchy of
semiclassical propagators at different levels of quantum accuracy.
In the past, several groups have beaten this path in the search for
a hybrid semiclassical propagator. For example, Zhang and Pollak,
in their SC-IVR perturbative series method \cite{Pollak_perturbationSeries},
treated the system variables with the Herman-Kluk (HK) prefactor \cite{Herman_Kluk_2}
and the bath variables using a prefactor free propagation. This is
obtained by forcing a unitary pre-exponential factor of the HK propagator,
which is expensive to compute as the dimensionality of the problem
increases \cite{Pollak_prefactorfree_05}. Earlier, Ovchinnikov and
Apkarian focused on condensed phase spectroscopy by using second-
and zeroth order approximations in stationary phase of the exact quantum
propagator, respectively the van Vleck and a prefactor-free van Vleck
propagator in Initial Value Representation (vV-IVR) \cite{Apkarian}.
At about the same time, Sun and Miller introduced a mixed semiclassical-classical
model, where the vV-IVR is either used in first or zeroth order approximation,
according to the amount of quantum delocalization retained around
each classical trajectory \cite{MIller_Sun_mixedSCclassical}. Also
the Filinov smoothing can be used to tune the semiclassical propagator
\cite{Miller336_GeneralizedFBSCIVR_01}, as recently shown \cite{Nandini_mixed_quantum_classical}.
In 2006, one of us (FG) implemented a similar idea for the Gaussian
dressed semiclassical dynamics of the HK propagator \cite{Grossmann_hybrid_06}.
More specifically, Gaussian wave packet propagation with the HK propagator
is equivalent to a Thawed Gaussian wave packet dynamics (TGWD) \cite{Heller_thawedgaussian}
if the phase space integral is approximated to second order in the
exponent around the phase space center of the wave packet (linearization
of the classical trajatories) \cite{Grossmann_thwg,Deshpande_Ezra_2006-1}.
If the transition from HK to TGWD is performed analytically only for
a selected set of degrees of freedom, one obtains a semiclassical
hybrid dynamics (SCHD) in the same spirit as described above. The
HK propagator is quite accurate and definitely superior to a single
trajectory TGWD, but computationally much more expensive for many
degrees of freedom. However, the full HK propagator is not necessary
to describe the dynamics for harmonic like modes, where the TGWD is
already quite accurate\cite{Jiri_oligotiophenes_14,Jiri_ammonia_2015,Jiri_Miroslav_MolPHys_2012,Pollak_Conte_TGWD}.
The semiclassical hybrid propagation takes advantage of both methods
when the system is treated at the level of HK and the bath with the
TGWD.

For the pure HK propagator the method of time-averaging \cite{Kay-TA}
has been shown to improve the numerical efficiency for the calculation
of spectra, after the so-called separable approximation \cite{Alex_Mik}.
The goal in the following is to apply the time-averaging idea together
with the SCHD propagation scheme to produce a mixed semiclassical
time-averaging propagator for spectroscopic calculations.

The paper is organized in the following way. Section \ref{sec:Time-averaging-SC-IVR}
recalls the time-averaging semiclassical method. Section \ref{sec:Mixed}
introduces a new mixed semiclassical progapator and Section \ref{sec:Sep-Mixed}
presents a computationally cheap version of this propagator based
on a separable approximation. In Section \ref{sec:Results} results
for the Caldeira-Leggett model Hamiltonian are presented and compared
with exact quantum wave packet propagations. After a discussion of
the results in Section \ref{sec:Results}, Conclusions are drawn in
Section \ref{sec:Conclusions}.

\section{The time-averaging SC-IVR method for power spectra calculations\label{sec:Time-averaging-SC-IVR}}

This paper focuses on spectroscopic calculations. More specifically,
we want to calculate the power spectra components $I\left(E\right)$
of a given reference state $\left|\chi\right\rangle $ subject to
the Hamiltonian $\hat{H}$,
\begin{align}
\mbox{I}\left(E\right)= & \sum_{i}\left|\left\langle \chi\left|\right.\psi_{i}\right\rangle \right|^{2}\delta(\mbox{E}-\mbox{E}_{i}),\label{eq:spectrum1}
\end{align}
where $E_{i}$ are the eigen-energies that we are interested in and
$\left|\psi_{i}\right\rangle $ are the associated eigen-functions
of the Hamiltonian. By representing the Dirac-delta in terms of a
Fourier integral, Eq. (\ref{eq:spectrum1}) can be written as \cite{Heller_review_autocorrel}

\begin{align}
\mbox{I}(E)= & \frac{1}{2\pi\hbar}\int\limits _{-\infty}^{\infty}\text{d}t\:\text{e}^{\text{i}Et/\hbar}\left\langle \chi\left|\text{e}^{-\text{i}\hat{H}t/\hbar}\right|\chi\right\rangle \label{eq:spectrum2}
\end{align}
In semiclassical dynamics, the time evolution of Eq. (\ref{eq:spectrum2})
can be calculated using the HK propagator\textcolor{black}{
\begin{eqnarray}
\text{e}^{-\textmd{i}\hat{H}t/\hbar} & = & \frac{1}{\left(2\pi\hbar\right)^{F}}\int\text{d}\mathbf{p}\left(0\right)\int\text{d}\mathbf{q}\left(0\right)\: C_{t}\left(\mathbf{p}(0),\mathbf{q}\left(0\right)\right)\label{eq:SCTime_evolution}\\
 &  & \times\text{e}^{\text{i}S_{t}\left(\mathbf{p}\left(0\right),\mathbf{q}\left(0\right)\right)/\hbar}\left|\mathbf{p}\left(t\right),\mathbf{q}\left(t\right)\left\rangle \right\langle \mathbf{p}\left(0\right),\mathbf{q}\left(0\right)\right|,\nonumber
\end{eqnarray}
where $\left(\mathbf{p}\left(t\right),\mathbf{q}\left(t\right)\right)$
is the set of }$2F-$dimensional\textcolor{black}{{} classically-evolved
phase space coordinates and $S_{t}$ is the corresponding classical
action. The integral representation of the propagator goes back to
the Frozen Gaussian Approximation of Heller \cite{Heller_frozengaussian}.
The pre-exponential factor
\begin{eqnarray}
C_{t}\left(\mathbf{p}\left(0\right),\mathbf{q}\left(0\right)\right) & = & \sqrt{\det\left[\frac{1}{2}\left(\frac{\partial\mathbf{q}\left(t\right)}{\partial\mathbf{q}\left(0\right)}+\frac{\partial\mathbf{p}\left(t\right)}{\partial\mathbf{p}\left(0\right)}-\text{i}\hbar\gamma\frac{\partial\mathbf{q}\left(t\right)}{\partial\mathbf{p}\left(0\right)}+\frac{\text{i}}{\gamma\hbar}\frac{\partial\mathbf{p}\left(t\right)}{\partial\mathbf{q}\left(0\right)}\right)\right]}\label{eq:prefactor}
\end{eqnarray}
takes into account second-order quantum delocalizations about the
classical paths \cite{Miller_avd_74,Kay_94,Manolopoulos,Coker,Takatsuka_eigenstates_05,Roy,Pollak_autocorrelation,Miller_vari,Miller_Tao_timedependentsampling_11,Roy_AbinitioSCIVR,Maitra_SCmaps_00_filter,Liu_IJQC_review_15,Tao_nonadaJPCA_13,Burant_Batista}.
The prefactor in Eq. (\ref{eq:prefactor}) is obtained if the initial
state is represented in the coherent-state basis set \cite{Herman_Kluk_2,Grossmann-Xavier},
which, for many degrees of freedom, is gi}ven by the direct product
of one dimensional coherent states\textcolor{black}{
\begin{align}
\left\langle \mathbf{q}\left|\right.\mathbf{p}\left(t\right),\mathbf{q}\left(t\right)\right\rangle  & =\prod_{i}\left(\frac{\mathbf{\gamma}_{i}}{\pi}\right)^{F/4}\mbox{\ensuremath{\exp}}\left[-\frac{\gamma_{i}}{2}\left(q_{i}-q_{i}\left(t\right)\right)^{2}+\frac{\text{i}}{\hbar}p_{i}\left(t\right)\left(q_{i}-q_{i}\left(t\right)\right)\right],\label{eq:coherent_state}
\end{align}
where $\gamma_{i}$ is fixed. For spectroscopic calculations, $\gamma_{i}$
is conveniently set equal to the width of the harmonic oscillator
approximation to the vibrational wave function for the i-th normal
mode. }

\textcolor{black}{To accelerate the Monte Carlo phase space integration
of Eq. (\ref{eq:SCTime_evolution}), a time-averaging (TA) integral
has been introduced \cite{Kay-TA,Alex_Mik,Ceotto_MCSCIVR}. Then,
Eq. (\ref{eq:spectrum2}) becomes
\begin{eqnarray}
I\left(E\right) & = & \frac{1}{\left(2\pi\hbar\right)^{F}}\int\text{d}\mathbf{p}\left(0\right)\int\text{d}\mathbf{q}\left(0\right)\frac{\mbox{Re}}{\pi\hbar T}\int_{0}^{T}\text{d}t_{1}\int_{t_{1}}^{T}\text{d}t_{2}\: C_{t_{2}}\left(\mathbf{p}\left(t_{1}\right),\mathbf{q}\left(t_{1}\right)\right)\label{eq:TA_spectrdens}\\
 &  & \times\left\langle \chi\left|\right.\mathbf{p}\left(t_{2}\right),\mathbf{q}\left(t_{2}\right)\right\rangle \text{e}^{\text{i}\left(S_{t_{2}}\left(\mathbf{p}\left(0\right),\mathbf{q}\left(0\right)\right)+Et_{2}\right)/\hbar}\left[\left\langle \chi\left|\right.\mathbf{p}\left(t_{1}\right),\mathbf{q}\left(t_{1}\right)\right\rangle \text{e}^{\text{i}\left(S_{t_{1}}\left(\mathbf{p}\left(0\right),\mathbf{q}\left(0\right)\right)+Et_{1}\right)/\hbar}\right]^{*},\nonumber
\end{eqnarray}
where the initial phase space point $\left(\mathbf{p}\left(0\right),\mathbf{q}\left(0\right)\right)$
has been evolved to the Fourier integration time $t_{1}$, respectively
to the time averaging time $t_{2}$, and $T$ is the total simulation
time. The computational effort of the double time variable integration
can be alleviated by approximating the pre-exponential factor as $C_{t_{2}}\left(\mathbf{p}\left(t_{1}\right),\mathbf{q}\left(t_{1}\right)\right)\thickapprox\exp\left[\text{i}\left(\phi\left(t_{2}\right)-\phi\left(t_{1}\right)\right)/\hbar\right]$,
where $\phi\left(t\right)/\hbar=\mbox{phase}\left[C_{t}\left(\mathbf{p}\left(0\right),\mathbf{q}\left(0\right)\right)\right]$.
This is called the ``separable approximation'' and Eq. (\ref{eq:TA_spectrdens})
becomes
\begin{eqnarray}
I\left(E\right) & = & \frac{1}{\left(2\pi\hbar\right)^{F}}\frac{1}{2\pi\hbar T}\int\text{d}\mathbf{p}\left(0\right)\int\text{d}\mathbf{q}\left(0\right)\label{eq:sep_approx}\\
 &  & \times\left|\int_{0}^{T}\text{d}t\left\langle \chi\left|\right.\mathbf{p}\left(t\right),\mathbf{q}\left(t\right)\right\rangle \text{e}^{\text{i}\left[S_{t}\left(\mathbf{p}\left(0\right),\mathbf{q}\left(0\right)\right)+Et+\phi_{t}\left(\mathbf{p}\left(0\right),\mathbf{q}\left(0\right)\right)\right]/\hbar}\right|^{2}\nonumber
\end{eqnarray}
in the separable approximation. The integral in Eq. (\ref{eq:sep_approx})
contains a single and positive-definite time integrand, which is more
stable numerically than that one of Eq. (\ref{eq:TA_spectrdens}).
Furthermore, the numerical evaluation of (\ref{eq:sep_approx}) is
quite accurate\cite{Alex_Mik,Ceotto_MCSCIVR} and much less computationally
demanding than that of (\ref{eq:TA_spectrdens}). }

\textcolor{black}{A recent implementation of Eqs. (\ref{eq:TA_spectrdens})
and (\ref{eq:sep_approx}) is that of Ceotto et al., called ``Multiple
Coherent SC-IVR'' (MC-SC-IVR)} \cite{Ceotto_1traj,Ceotto_MCSCIVR,Ceotto_eigenfunctions,Ceotto_cursofdimensionality_11,Ceotto_david,Ceotto_acceleratedSCIVR,Ceotto_NH3,Ceotto_Zhang_JCTC},\textcolor{black}{{}
where the reference state }$\left|\chi\right\rangle =\sum_{i=1}^{N_{\text{states}}}\left|\mathbf{p}_{\text{eq}}^{i},\mathbf{q}_{\text{eq}}^{i}\right\rangle $
is written as a combination of coherent states placed nearby the classical
phase space points $\left(\mathbf{p}_{\text{eq}}^{i},\mathbf{q}_{\text{eq}}^{i}\right)$,
where $\mathbf{q}_{\text{eq}}^{i}$ is an equilibrium position and
$\mathbf{p}_{\text{eq}}^{i}$ corresponds, in a harmonic fashion,
to excited vibrational states, i.e. $\left(p_{i,\text{eq}}\right)^{2}/2m=\hbar\omega_{i}\left(n+1/2\right)$.
In this way, one can reduce the number of trajectories to a few ``eigen-trajectories'',
one for each coherent state location, corresponding to the harmonic
sequence of eigenvalues. The Gaussian delocalization showed to alleviate
the shortcomings of a global harmonic approximation, that one would
perform if just a single trajectory was used, and to fully provide
anharmonic effects. MC-SC-IVR has been successfully applied to gas
phase spectra calculations of the $\mbox{H}_{2}\mbox{O}$ molecule
\cite{Ceotto_MCSCIVR}, and $\mbox{CO}$ molecules chemisorbed on
a $\mbox{Cu}(100)$ surface using a pre-computed potential \cite{Ceotto_david}.
Using a direct \emph{ab initio} approach, vibrational energies for
$\mbox{CO}{}_{2}$ \cite{Ceotto_MCSCIVR,Ceotto_1traj}, $\mbox{H}{}_{2}\mbox{CO}$
\cite{Ceotto_cursofdimensionality_11} and the ammonia umbrella inversion
\cite{Ceotto_NH3}, as well as $\mbox{CO}{}_{2}$ vibrational eigenfunctions
\cite{Ceotto_eigenfunctions}, have been calculated. The MC-SC-IVR
computational time is dramatically reduced when the method is implemented
for GPU architectures \cite{Ceotto_GPU}.

\section{A mixed semiclassical power spectrum method\label{sec:Mixed}}

The idea of the SCHD is based on a mixed semiclassical propagator
approach. We partition the $2F$ phase space variables into $2F_{\text{hk}}$
for the system phase space and $2F_{\text{tg}}$ for the bath phase
space. The HK level of accuracy is reserved for the system only, indicated
by the subscript hk, whereas the bath phase space variables are treated
on the thawed Gaussian level (subscript tg). The reference state is
chosen as $\left|\chi\right\rangle =\left|\mathbf{p}_{\text{eq}}\left(0\right),\mathbf{q}_{\text{eq}}\left(0\right)\right\rangle $
as explained above, and the initial phase space coordinate vectors
are subdivided as
\begin{equation}
\mathbf{p}_{\text{eq}}\left(0\right)\equiv\left(\begin{array}{c}
\mathbf{p}_{\text{eq},\:\text{hk}}\left(0\right)\\
\mathbf{p}_{\text{eq},\:\text{tg}}\left(0\right)
\end{array}\right);\;\;\;\mathbf{q}_{\text{eq}}\left(0\right)\equiv\left(\begin{array}{c}
\mathbf{q}_{\text{eq},\:\text{hk}}\left(0\right)\\
\mathbf{q}_{\text{eq},\:\text{tg}}\left(0\right)
\end{array}\right)\label{eq:initial_mixed_coord}
\end{equation}
where the bath starting coordinates are always assumed to be at the
equilibrium positions. This phase space variable partitioning is motivated
by considering that the TGWD exactly reproduces the full harmonic
spectrum, as shown in Appendix \ref{sec:Appendix-A:TWD}. The partitioning
should be well suited for a harmonic-like motion of the bath degrees
of freedom.

Following \cite{Grossmann_hybrid_06}, we approximate the evolution
of the phase space coordinates in (\ref{eq:initial_mixed_coord})
at each time step as
\begin{eqnarray}
\mathbf{q}(t)\equiv\left(\begin{matrix}\mathbf{q}_{\text{hk}}\left(t\right)\\
\mathbf{q}_{\text{tg}}\left(t\right)
\end{matrix}\right) & = & \mathbf{q}_{\text{eq}}\left(t\right)+\mathbf{m}_{22}\left(t\right)\delta\mathbf{q}_{\text{tg}}+\mathbf{m}_{21}\left(t\right)\delta\mathbf{p}_{\text{tg}}\label{eq:HYB-qt}\\
\mathbf{p}(t)\equiv\left(\begin{matrix}\mathbf{p}_{\text{hk}}\left(t\right)\\
\mathbf{p}_{\text{tg}}\left(t\right)
\end{matrix}\right) & = & \mathbf{p}_{\text{eq}}\left(t\right)+\mathbf{m}_{12}\left(t\right)\delta\mathbf{q}_{\text{tg}}+\mathbf{m}_{11}\left(t\right)\delta\mathbf{p}_{\text{tg}}\label{eq:HYB-pt}
\end{eqnarray}
where the trajectory coordinates are linearly expanded for the bath
DOFs only. The matrices
\begin{equation}
\begin{array}{c}
\mathbf{m}_{11}\left(t\right)=\frac{\partial\mathbf{p}_{\text{eq},}\left(t\right)}{\partial\mathbf{p}_{\text{eq},\:\text{tg}}\left(0\right)};\;\;\;\mathbf{m}_{12}\left(t\right)=\frac{\partial\mathbf{p}_{\text{eq},}\left(t\right)}{\partial\mathbf{q}_{\text{eq},\:\text{tg}}\left(0\right)};\\
\mathbf{m}_{21}\left(t\right)=\frac{\partial\mathbf{q}_{\text{eq},}\left(t\right)}{\partial\mathbf{p}_{\text{eq},\:\text{tg}}\left(0\right)};\;\;\;\mathbf{m}_{22}\left(t\right)=\frac{\partial\mathbf{q}_{\text{eq}}\left(t\right)}{\partial\mathbf{q}_{\text{eq},\:\text{tg}}\left(0\right)};
\end{array}\label{eq:HYB-mij-1}
\end{equation}
are non-square $F\times F_{\text{tg}}$ dimensional and the displacements
\begin{equation}
\begin{array}{c}
\delta\mathbf{p}_{\text{tg}}=\mathbf{p}_{\text{tg}}\left(0\right)-\mathbf{p}_{\text{eq},\:\text{tg}}\left(0\right)\\
\delta\mathbf{q}_{\text{tg}}=\mathbf{q}_{\text{tg}}\left(0\right)-\mathbf{q}_{\text{eq},\:\text{tg}}\left(0\right)
\end{array}\label{eq:displac_vec}
\end{equation}
are $F_{\text{tg}}$ dimensional. To apply this approximation to Eq.
(\ref{eq:sep_approx}), we express it as\textcolor{black}{
\begin{eqnarray}
I\left(E\right) & = & \frac{1}{\left(2\pi\hbar\right)^{F}}\frac{1}{2\pi\hbar T}\int\text{d}\mathbf{p}\left(0\right)\int\text{d}\mathbf{q}\left(0\right)\nonumber \\
 &  & \times\left|\int_{0}^{T}\text{d}t\:\text{e}^{\text{i}\left(S_{t}\left(\mathbf{p}\left(0\right),\mathbf{q}\left(0\right)\right)+Et+\phi_{t}\left(\mathbf{p}\left(0\right),\mathbf{q}\left(0\right)\right)/\hbar\right)}\right.\label{eq:sep_approx-2}\\
 &  & \times\exp\left\{ -\frac{1}{4}\left(\mathbf{q}\left(t\right)-\mathbf{q}_{\text{eq}}\left(0\right)\right)^{\text{T}}\boldsymbol{\gamma}\left(\mathbf{q}\left(t\right)-\mathbf{q}_{\text{eq}}\left(0\right)\right)\right\} \nonumber \\
 &  & \times\exp\left\{ -\frac{1}{4\hbar^{2}}\left(\mathbf{p}\left(t\right)-\mathbf{p}_{\text{eq}}\left(t\right)\right)^{\text{T}}\mathbf{\boldsymbol{\gamma}}^{-1}\left(\mathbf{p}\left(t\right)-\mathbf{p}_{\text{eq}}\left(0\right)\right)\right\} \nonumber \\
 &  & \times\left.\exp\left\{ +\frac{\text{i}}{2\hbar}\left(\mathbf{q}_{\text{eq}}\left(0\right)-\mathbf{q}\left(t\right)\right)^{\text{T}}\left(\mathbf{p}\left(t\right)+\mathbf{p}_{\text{eq}}\left(0\right)\right)\right\} \right|^{2}\nonumber
\end{eqnarray}
where the coherent reference state is explicitly written out.}

We now express all quantities appearing in (\ref{eq:sep_approx-2})
in terms of the trajectory in Eqs. (\ref{eq:HYB-qt}) and (\ref{eq:HYB-pt}).
The classical action becomes
\begin{eqnarray}
S_{t}\left(\mathbf{p}\left(0\right),\mathbf{q}\left(0\right)\right) & = & S_{t}\left(\mathbf{p}_{\text{hk}}\left(0\right),\mathbf{q}_{\text{hk}}\left(0\right),\mathbf{p}_{\text{eq},\:\text{tg}}\left(0\right),\mathbf{q}_{\text{eq},\:\text{tg}}\left(0\right)\right)\label{eq:HYB-action-1}\\
 &  & +\mathbf{p}_{\text{eq}}^{\text{T}}(t)\mathbf{m}_{21}\left(t\right)\delta\mathbf{p}_{\text{tg}}+\left(\mathbf{p}_{\text{eq}}^{\text{T}}(t)\mathbf{m}_{22}\left(t\right)-\mathbf{p}_{\text{eq},\:0,\:\text{tg}}^{\text{T}}\right)\delta\mathbf{q}_{\text{tg}}\nonumber \\
 &  & +\frac{1}{2}\delta\mathbf{p}_{\text{tg}}^{\text{T}}\mathbf{m}_{11}^{\text{T}}\left(t\right)\mathbf{m}_{21}\left(t\right)\delta\mathbf{p}_{\text{tg}}+\frac{1}{2}\delta\mathbf{q}_{\text{tg}}^{\text{T}}\mathbf{m}_{12}^{\text{T}}\left(t\right)\mathbf{m}_{22}\left(t\right)\delta\mathbf{q}_{\text{tg}}\nonumber \\
 &  & +\delta\mathbf{q}_{\text{tg}}^{\text{T}}\mathbf{m}_{12}^{\text{T}}\left(t\right)\mathbf{m}_{21}\left(t\right)\delta\mathbf{p}_{\text{tg}}\nonumber
\end{eqnarray}
up to the second order in fluctuations for the bath subspace\cite{Grossmann_hybrid_06}.
In the same fashion, we insert (\ref{eq:HYB-qt}) and (\ref{eq:HYB-pt})
into the coherent states overlap, retain the terms up to the second
order and obtain three approximated exponential terms. By inserting
these terms and Eq. (\ref{eq:HYB-action-1}) into the power spectrum
expression (\ref{eq:sep_approx-2}), we obtain the mixed semiclassical
power spectrum approximation\textcolor{black}{
\begin{eqnarray}
I\left(E\right) & = & \frac{1}{\left(2\pi\hbar\right)^{F}}\frac{1}{2\pi\hbar T}\int\text{d}\mathbf{p}\left(0\right)\int\text{d}\mathbf{q}\left(0\right)\left|\int_{0}^{T}\text{d}t\:\text{e}^{\text{i}\left[Et+\phi_{t}\left(\mathbf{p}\left(0\right),\mathbf{q}\left(0\right)\right)\right]/\hbar}\right.\label{eq:sep_approx-3}\\
 &  & \times\exp\left\{ -\frac{1}{4}\left(\mathbf{q}_{\text{hk}}\left(t\right)-\mathbf{q}_{\text{hk}}\left(0\right)\right)^{\text{T}}\mathbf{\boldsymbol{\gamma}_{\text{hk}}}\left(\mathbf{q}_{\text{hk}}\left(t\right)-\mathbf{q}_{hk}\left(0\right)\right)\right\} \nonumber \\
 &  & \times\exp\left\{ -\frac{1}{4\hbar^{2}}\left(\mathbf{p}_{\text{hk}}\left(t\right)-\mathbf{p}_{\text{hk}}\left(0\right)\right)^{\text{T}}\mathbf{\boldsymbol{\gamma}}_{\text{hk}}^{-1}\left(\mathbf{p}_{\text{hk}}\left(t\right)-\mathbf{p}_{\text{hk}}\left(0\right)\right)\right\} \nonumber \\
 &  & \times\exp\left\{ +\frac{\text{i}}{2\hbar}\left(\mathbf{q}_{\text{hk}}\left(0\right)-\mathbf{q}_{\text{hk}}\left(t\right)\right)^{\text{T}}\left(\mathbf{p}_{\text{hk}}\left(t\right)+\mathbf{p}_{\text{hk}}\left(0\right)\right)\right\} \nonumber \\
 &  & \times\left.\mbox{exp}\left\{ -\left(\begin{array}{c}
\delta\mathbf{p}_{\text{tg}}\\
\delta\mathbf{q}_{\text{tg}}
\end{array}\right)^{\text{T}}\mathbf{A}\left(t\right)\left(\begin{array}{c}
\delta\mathbf{p}_{\text{tg}}\\
\delta\mathbf{q}_{\text{tg}}
\end{array}\right)+\mathbf{b}^{\text{T}}\left(\begin{array}{c}
\delta\mathbf{p}_{\text{tg}}\\
\delta\mathbf{q}_{\text{tg}}
\end{array}\right)+c_{t}\right\} \right|^{2},\nonumber
\end{eqnarray}
}where we have introduced the $F_{\text{tg}}\times F_{\text{tg}}$
diagonal matrices $\boldsymbol{\gamma}_{\text{tg}}$ and the $F_{\text{hk}}\times F_{\text{hk}}$
diagonal matrices $\boldsymbol{\gamma}_{\text{hk}}$ comprising the
respective width parameters. In the last exponential of (\ref{eq:sep_approx-3}),
the terms have been collected according to the respective power of
$\delta\mathbf{p}_{\text{tg}}$ and $\delta\mathbf{q}_{\text{tg}}$.
The zeroth order terms are
\begin{eqnarray}
c_{t} & = & \frac{\text{i}}{\hbar}S_{t}\left(\mathbf{p}_{\text{hk}}\left(0\right),\mathbf{q}_{\text{hk}}\left(0\right),\mathbf{p}_{\text{eq},\:\text{tg}}\left(0\right),\mathbf{q}_{\text{eq},\:\text{tg}}\left(0\right)\right)\label{eq:c_t}\\
 &  & -\frac{1}{4}\left(\mathbf{q}_{\text{eq},\:\text{tg}}\left(t\right)-\mathbf{q}_{\text{eq},\:\text{tg}}\left(0\right)\right)^{\text{T}}\mathbf{\boldsymbol{\gamma}}_{\text{tg}}\left(\mathbf{q}_{\text{eq},\:\text{tg}}\left(t\right)-\mathbf{q}_{\text{eq},\:\text{tg}}\left(0\right)\right)\nonumber \\
 &  & -\frac{1}{4\hbar^{2}}\left(\mathbf{p}_{\text{eq},\:\text{tg}}\left(t\right)-\mathbf{p}_{\text{eq},\:\text{tg}}\left(0\right)\right)^{\text{T}}\mathbf{\boldsymbol{\gamma}}_{\text{tg}}^{-1}\left(\mathbf{p}_{\text{eq},\:\text{tg}}\left(t\right)-\mathbf{p}_{\text{eq},\:\text{tg}}\left(0\right)\right)\nonumber \\
 &  & +\frac{\text{i}}{2\hbar}\left(\mathbf{q}_{\text{eq},\:\text{tg}}\left(0\right)-\mathbf{q}_{\text{eq},\:\text{tg}}\left(t\right)\right)^{\text{T}}\left(\mathbf{p}_{\text{eq},\:\text{tg}}\left(t\right)+\mathbf{p}_{\text{eq},\:\text{tg}}\left(0\right)\right)\nonumber
\end{eqnarray}
 and the coefficients of the second order terms are collected in the
matrix $\mathbf{A}\left(t\right)$ composed of the following $F_{\text{tg}}\times F_{\text{tg}}$
blocks
\begin{eqnarray}
\mathbf{A}_{11}\left(t\right) & = & \frac{1}{4}\mathbf{m}_{21}^{\text{T}}\left(t\right)\mathbf{\boldsymbol{\gamma}}\mathbf{m}_{21}\left(t\right)+\frac{1}{4\hbar^{2}}\mathbf{m}_{11}^{\text{T}}\left(t\right)\mathbf{\boldsymbol{\gamma}}^{-1}\mathbf{m}_{11}\left(t\right)\label{eq:A_matrix}\\
\mathbf{A}_{12}\left(t\right) & = & \frac{1}{4}\mathbf{m}_{21}^{\text{T}}\left(t\right)\boldsymbol{\gamma}\mathbf{m}_{22}\left(t\right)+\frac{1}{4\hbar^{2}}\mathbf{m}_{11}^{\text{T}}\left(t\right)\boldsymbol{\gamma}^{-1}\mathbf{m}_{12}\left(t\right)\nonumber \\
\mathbf{A}_{21}\left(t\right) & = & \frac{1}{4}\mathbf{m}_{22}^{\text{T}}\left(t\right)\boldsymbol{\gamma}\mathbf{m}_{21}\left(t\right)+\frac{1}{4\hbar^{2}}\mathbf{m}_{12}^{\text{T}}\left(t\right)\boldsymbol{\gamma}^{-1}\mathbf{m}_{11}\left(t\right)+\frac{\text{i}}{2\hbar}\nonumber \\
\mathbf{A}_{22}\left(t\right) & = & \frac{1}{4}\mathbf{m}_{22}^{\text{T}}\left(t\right)\boldsymbol{\gamma}\mathbf{m}_{22}\left(t\right)+\frac{1}{4\hbar^{2}}\mathbf{m}_{12}^{\text{T}}\left(t\right)\mathbf{\boldsymbol{\gamma}}^{-1}\mathbf{m}_{12}\left(t\right).\nonumber
\end{eqnarray}
The coefficients of the first order terms in $\delta\mathbf{p}_{\text{tg}}$
and $\delta\mathbf{q}_{\text{tg}}$ are collected in a $2F_{\text{tg}}$
dimensional vector of the type
\begin{equation}
\mathbf{b}_{t}\equiv\left(\begin{array}{c}
\mathbf{b}_{1,t}\\
\mathbf{b}_{2,t}
\end{array}\right)\label{eq:B_definition}
\end{equation}
where
\begin{eqnarray}
\mathbf{b}_{1,t}^{\text{T}} & = & -\frac{1}{2}\left(\mathbf{q}\left(t\right)-\mathbf{q}\left(0\right)\right)^{\text{T}}\left[\mathbf{\boldsymbol{\gamma}}\mathbf{m}_{21}\left(t\right)+\frac{\text{i}}{\hbar}\mathbf{m}_{11}\left(t\right)\right]\label{eq:b1}\\
 &  & -\frac{1}{2\hbar^{2}}\left(\mathbf{p}\left(t\right)-\mathbf{p}\left(0\right)\right)^{\text{T}}\left[\mathbf{\boldsymbol{\gamma}}^{-1}\mathbf{m}_{11}\left(t\right)-\text{i}\hbar\mathbf{m}_{21}\left(t\right)\right]\nonumber \\
\mathbf{b}_{2,t}^{\text{T}} & = & -\frac{1}{2}\left(\mathbf{q}\left(t\right)-\mathbf{q}\left(0\right)\right)^{\text{T}}\left[\mathbf{\boldsymbol{\gamma}}\mathbf{m}_{22}\left(t\right)+\frac{\text{i}}{\hbar}\mathbf{m}_{12}\left(t\right)\right]\label{eq:b2}\\
 &  & -\frac{1}{2\hbar^{2}}\left(\mathbf{p}\left(t\right)-\mathbf{p}\left(0\right)\right)^{\text{T}}\left[\mathbf{\boldsymbol{\gamma}}^{-1}\mathbf{m}_{12}\left(t\right)-\text{i}\hbar\mathbf{m}_{22}\left(t\right)\right]-\frac{\text{i}}{\hbar}\mathbf{p}_{\text{eq},\:\text{tg}}^{\text{T}}.\nonumber
\end{eqnarray}
To carry out the Gaussian integration in $\delta\mathbf{p}_{\text{tg}}$
and $\delta\mathbf{q}_{\text{tg}}$ in (\ref{eq:sep_approx-3}), we
first unravel the modulus squared, then change the coordinates in
the bath subspace from the phase space ones to the displacement ones
of (\ref{eq:displac_vec}) and obtain\textcolor{black}{
\begin{eqnarray}
I\left(E\right) & = & \frac{1}{\left(2\pi\hbar\right)^{F}}\frac{\text{Re}}{\pi\hbar T}\int_{0}^{T}\text{d}t_{1}\int_{t_{1}}^{T}\text{d}t_{2}\int\text{d}\mathbf{p}_{\text{hk}}\left(0\right)\int\text{d}\mathbf{q}_{\text{hk}}\left(0\right)\label{eq:sep_approx-4}\\
 &  & \times\text{e}^{\text{i}\left[E\left(t_{1}-t_{2}\right)+\phi_{t_{1}}\left(\mathbf{p}\left(0\right),\mathbf{q}\left(0\right)\right)-\phi_{t_{2}}\left(\mathbf{p}\left(0\right),\mathbf{q}\left(0\right)\right)\right]/\hbar}\nonumber \\
 &  & \times\left\langle \mathbf{p}_{\text{eq},\:\text{hk}}\left(0\right),\mathbf{q}_{\text{eq},\:\text{hk}}\left(0\right)\left|\right.\mathbf{p}\left(t_{1}\right),\mathbf{q}\left(t_{1}\right)\right\rangle \left\langle \mathbf{p}\left(t_{2}\right),\mathbf{q}\left(t_{2}\right)\left|\right.\mathbf{p}_{\text{eq},\:\text{hk}}\left(0\right),\mathbf{q}_{\text{eq},\:\text{hk}}\left(0\right)\right\rangle \nonumber \\
 &  & \times\int\text{d}\delta\mathbf{p}_{\text{tg}}\left(0\right)\int\text{d}\delta\mathbf{q}_{\text{tg}}\left(0\right)\mbox{exp}\left\{ -\left(\begin{array}{c}
\delta\mathbf{p}_{\text{tg}}\\
\delta\mathbf{q}_{\text{tg}}
\end{array}\right)^{\text{T}}\left(\mathbf{A}\left(t_{1}\right)+\mathbf{A}^{*}\left(t_{2}\right)\right)\left(\begin{array}{c}
\delta\mathbf{p}_{\text{tg}}\\
\delta\mathbf{q}_{\text{tg}}
\end{array}\right)\right\} \nonumber \\
 &  & \times\mbox{exp}\left\{ \left(\mathbf{b}_{t_{1}}+\mathbf{b}_{t_{2}}^{*}\right)^{\text{T}}\left(\begin{array}{c}
\delta\mathbf{p}_{\text{tg}}\\
\delta\mathbf{q}_{\text{tg}}
\end{array}\right)+c_{t_{1}}+c_{t_{2}}^{*}\right\} .\nonumber
\end{eqnarray}
Finally, we integrate over the bath displacements coordinates and
obtain the present mixed semiclassical approximation of (\ref{eq:TA_spectrdens})
\begin{eqnarray}
I\left(E\right) & = & \frac{1}{\left(2\hbar\right)^{F}}\frac{1}{\pi^{F_{\text{hk}}}}\frac{\text{Re}}{\pi\hbar T}\int\text{d}\mathbf{p}_{\text{hk}}\left(0\right)\int\text{d}\mathbf{q}_{\text{hk}}\left(0\right)\int_{0}^{T}\text{d}t_{1}\int_{t_{1}}^{T}\text{d}t_{2}\label{eq:sep_approx-5}\\
 &  & \times\text{e}^{\text{i}\left[E\left(t_{1}-t_{2}\right)+\phi_{t_{1}}\left(\mathbf{p}\left(0\right),\mathbf{q}\left(0\right)\right)-\phi_{t_{2}}\left(\mathbf{p}\left(0\right),\mathbf{q}\left(0\right)\right)\right]/\hbar}\nonumber \\
 &  & \times\left\langle \mathbf{p}_{\text{eq},\:\text{hk}}\left(0\right),\mathbf{q}_{\text{eq},\:\text{hk}}\left(0\right)\left|\right.\mathbf{p}\left(t_{1}\right),\mathbf{q}\left(t_{1}\right)\right\rangle \left\langle \mathbf{p}\left(t_{2}\right),\mathbf{q}\left(t_{2}\right)\left|\right.\mathbf{p}_{\text{eq},\:\text{hk}}\left(0\right),\mathbf{q}_{\text{eq},\:\text{hk}}\left(0\right)\right\rangle \nonumber \\
 &  & \times\sqrt{\frac{1}{\det\left(\mathbf{A}\left(t_{1}\right)+\mathbf{A}^{*}\left(t_{2}\right)\right)}}\nonumber \\
 &  & \times\mbox{exp}\left\{ \frac{1}{4}\left(\mathbf{b}_{t_{1}}+\mathbf{b}_{t_{2}}^{*}\right)^{\text{T}}\left(\mathbf{A}\left(t_{1}\right)+\mathbf{A}^{*}\left(t_{2}\right)\right)^{-1}\left(\mathbf{b}_{t_{1}}+\mathbf{b}_{t_{2}}^{*}\right)+c_{t_{1}}+c_{t_{2}}^{*}\right\} .\nonumber
\end{eqnarray}
Eq. (\ref{eq:sep_approx-5}) keeps the system quantum evolution at
the level of accuracy of a semiclassical Herman-Kluk simulation implying
a Monte Carlo sampling over all the system coordinates and momenta.
Instead, the bath coordinates and momenta are not sampled, and all
initial conditions for the bath subspace trajectories are fixed at
equilibrium. However, quantum effects are included for the bath dynamics
by the thawed Gaussian quantum delocalization. We stress that only
the semiclassical propagator is approximated by using the hybrid idea.
The evolution of the underlying classical trajectories is still full
dimensional and system and bath are naturally coupled. It should also
be mentioned that due to the imaginary part of $\mathbf{A}(t)$ being
time independent, the radicand of the square root in (\ref{eq:sep_approx-5})
is real, so no extra care has to be taken in computing this quantity.}

\section{Separable approximations for the mixed semiclassical power spectrum
method\label{sec:Sep-Mixed}}

The double time integration of (\ref{eq:sep_approx-5}) can be quite
computationally demanding and the advantage of the approximated thawed
Gaussian dynamics for the bath coordinates is diminished by this double
integration. To recover a separable approximation of the type of the
original time-averaging SC-IVR expression of (\ref{eq:sep_approx})
for the mixed semiclassical expression of Eq. (\ref{eq:sep_approx-5}),
we approximate the exponential part as follows
\begin{eqnarray}
 &  & \frac{1}{4}\left(\mathbf{b}_{t_{1}}+\mathbf{b}_{t_{2}}^{*}\right)^{\text{T}}\left(\mathbf{A}\left(t_{1}\right)+\mathbf{A}^{*}\left(t_{2}\right)\right)^{-1}\left(\mathbf{b}_{t_{1}}+\mathbf{b}_{t_{2}}^{*}\right)\label{eq:separableB}\\
 &  & \approx\frac{1}{4}\mathbf{b}_{\text{m},t_{1}}^{\text{T}}\left(\mathbf{A}\left(t_{1}\right)+\mathbf{A}^{*}\left(t_{1}\right)\right)^{-1}\mathbf{b}_{\text{m},t_{1}}+\frac{1}{4}\left[\mathbf{b}_{\text{m},t_{2}}^{\text{T}}\left(\mathbf{A}\left(t_{2}\right)+\mathbf{A}^{*}\left(t_{2}\right)\right)^{-1}\mathbf{b}_{\text{m},t_{2}}\right]^{*},\nonumber
\end{eqnarray}
while the other exponential terms are naturally separable. The modified
vector $\mathbf{b}_{\mbox{m},t}$ is defined as
\begin{equation}
\mathbf{b}_{\text{m},t}=\left(\begin{array}{c}
\mathbf{b}_{1,t}\\
\mathbf{b}_{2,t}+\frac{\text{i}}{\hbar}\mathbf{p}_{\text{tg}}
\end{array}\right)\label{eq:modified_bt}
\end{equation}
The pre-exponential square root term is also not separable, and we
approximate it in the fashion of a geometric average by
\begin{equation}
\frac{1}{\sqrt{\det\left(\mathbf{A}\left(t_{1}\right)+\mathbf{A}^{*}\left(t_{2}\right)\right)}}\approx\left(\frac{1}{\det\left(\mathbf{A}\left(t_{1}\right)+\mathbf{A}^{*}\left(t_{1}\right)\right)}\right)^{1/4}\left(\frac{1}{\det\left(\mathbf{A}\left(t_{2}\right)+\mathbf{A}^{*}\left(t_{2}\right)\right)}\right)^{1/4}\label{eq:pre-exp_approx}
\end{equation}
Using Eq.s (\ref{eq:separableB}) and (\ref{eq:pre-exp_approx}),
the expression for the power spectrum is greatly simplified\textcolor{black}{
\begin{eqnarray}
I\left(E\right) & = & \frac{1}{\left(2\hbar\right)^{F}}\frac{1}{\pi^{F_{\text{hk}}}}\frac{1}{2\pi\hbar T}\int\text{d}\mathbf{p}_{\text{hk}}\left(0\right)\int\text{d}\mathbf{q}_{\text{hk}}\left(0\right)\left|\int_{0}^{T}\text{d}t\:\text{e}^{\text{i}\left[Et+\phi_{t}\left(\mathbf{p}\left(0\right),\mathbf{q}\left(0\right)\right)\right]/\hbar}\right.\label{eq:sep_B}\\
 &  & \times\left\langle \mathbf{p}_{\text{eq},\text{hk}}\left(0\right),\mathbf{q}_{\text{eq},\:\text{hk}}\left(0\right)\left|\right.\mathbf{p}\left(t\right),\mathbf{q}\left(t\right)\right\rangle \frac{1}{\left[\det\left(\mathbf{A}\left(t\right)+\mathbf{A}^{*}\left(t\right)\right)\right]^{1/4}}\nonumber \\
 &  & \times\left.\exp\left\{ \frac{1}{4}\mathbf{b}_{\text{m},t}^{\text{T}}\left(\mathbf{A}\left(t\right)+\mathbf{A}^{*}\left(t\right)\right)^{-1}\mathbf{b}_{\text{m},t}+c_{t}\right\} \right|^{2}\nonumber
\end{eqnarray}
}and much less computationally demanding since only a single time-integration
is now requested. Eq. (\ref{eq:sep_B}) still retains the full dimensional
classical evolution and the thawed Gaussian approximation for the
bath degrees of freedom only. However, the time-averaging part of
the thawed Gaussian dynamics is less accurate than that one in Eq.
(\ref{eq:sep_approx-5}).

Appendix A shows how the TGWD is exact for the power spectrum calculations
of the harmonic oscillator, including all vibrational levels. This
level of accuracy is preserved by our approximation. We demonstrate
that the peak positions of the spectrum are indeed reproduced correctly
in Appendix B. This is very important since the harmonic contribution
is often the main contribution for the potential energy surface of
bound systems.

\section{Results and Discussion\label{sec:Results}}

To test the accuracy of the power spectrum expression of Eq. (\ref{eq:sep_B}),
we consider a model system of a Morse oscillator coupled bilinearly
(the Caldeira-Leggett (CL) model \cite{CaldeiraLeggett_1983}) to
one, respectively two harmonic oscillators. We intentionally keep
the number of the bath modes low, since in this way we have exact
quantum wavepacket results available for comparison. The CL Hamiltonian
in atomic units is
\begin{equation}
H=H_{\text{s}}+\sum_{i=1}^{F_{\text{bath}}}\left\{ \frac{p_{i}^{2}}{2}+\frac{1}{2}\left[\omega_{i}y_{i}+\frac{c_{i}}{\omega_{i}}\left(s-s_{\text{eq}}\right)\right]^{2}\right\} \label{eq:Caldeira-Leggett_H}
\end{equation}
where $s$ is the system variable and $s_{\text{eq}}$ its equilibrium
position, and $y_{\text{i}}$ are the bath coordinates and the bath
masses are unitary. We choose bath parameters corresponding to an
Ohmic density with exponential cutoff, where the normalization factor
$c_{i}/\omega_{i}$ and a system-bath coupling strength $\eta$ are
defined as in \cite{Goletz_2009}.

The system Hamiltonian $H_{s}$ is that of a Morse potential
\begin{equation}
V_{\text{s}}\left(r\right)=D_{\text{e}}\left(1-\text{e}^{-\alpha\left(r-r_{\text{e}}\right)}\right)^{2}\label{eq:Morse_pot}
\end{equation}
where $D_{\text{e}}=0.057\:\mbox{a.u.}$, $r_{\text{e}}=0\:\mbox{a.u.}$,
and $\alpha=0.983\:\mbox{a.u.}.$The mass of the Morse oscillator
has been set to $M_{\text{r}}=1.165\times10^{5}\:\mbox{a.u.}$ and
the Morse frequency is $\omega_{\text{s}}=9.724\times10^{-4}\:\mbox{a.u.}$,
in order to reproduce the vibration of the $\mbox{I}_{2}$ molecule.

We will look at two different effective coupling strengths $\eta_{\text{eff}}=\eta/\left(m_{\text{s}}\omega_{\text{s}}\right)$
and three different cutoff frequencies $\omega_{\text{c}}$. Note
that $\omega_{\text{c}}$ is identical with the bath frequency for
the two dimensional case; in the three dimensional calculations there
is one additional bath oscillator of lower frequency. In one case,
we intentionally choose a bath frequency that is resonant with the
Morse potential's harmonic approximation, and another that is much
lower than $\omega_{\text{s}}$, and a third one in between. In the
resonant case, one might expect that the hybrid method is quite poor,
because the system, which is anharmonic, might drive the harmonic
bath into anharmonic dynamics, which is not accounted for by TGWD
part of the mixed semiclassical propagator.

The initial conditions are chosen in harmonic approximation as $(p\left(0\right),q\left(0\right))=\left(\sqrt{m_{\text{s}}\omega_{\text{s}}},0\right)$for
the system. Also the bath is initially at its equilibrium position
with harmonic zero point kinetic energy. We use $10^{4}$ trajectories
for the two-dimensional, and $5\times10^{4}$ trajectories for the
three-dimensional calculations. While this is enough to get the main
peak positions correctly, tight convergence of the hk result needs
more trajectories, as shown exemplary in Table \ref{tab:Number-of-trajectories}.
The time step is $\Delta t=T_{\text{s}}/20$ (where $T_{s}=2\pi/\omega_{\text{s}}$),
and the total number of (semiclassical) time steps is $2^{14}$, except
for Eq. (\ref{eq:sep_approx-5}) where we use only $2^{13}$ steps
because we have to do two time integrations in that case. The reference
quantum calculations are performed with the WavePacket software \cite{SchmidtLorenz2011}.

\begin{table}
\centering{}%
\begin{tabular}{l>{\centering}p{2cm}>{\centering}p{2cm}c}
 &  &  & \tabularnewline
\hline
\hline
\noalign{\vskip\doublerulesep}
 & TA HK sep (\ref{eq:sep_approx}) & TA mixed (\ref{eq:sep_approx-5}) & TA mixed-sep (\ref{eq:sep_B})\tabularnewline[\doublerulesep]
\hline
\noalign{\vskip\doublerulesep}
trajectories & $2\times10^{5}$ & $1\times10^{4}$ & $1\times10^{4}$\tabularnewline
\noalign{\vskip\doublerulesep}
time steps & $2^{14}$ & $2^{13}$ & $2^{14}$\tabularnewline
computational time & 10 hours & 33 hours & 40 min\tabularnewline
\hline
\hline
 &  &  & \tabularnewline
\end{tabular}\caption{Number of trajectories and computational times needed for tight convergence
of spectrum of a Morse oscillator coupled to 1 bath oscillator with
$\eta_{\text{eff}}=0.2$ and $\omega_{\text{bath}}=\omega_{\text{s}}/10$.
All propagation times from single CPU calculations on a standard desktop
computer.\label{tab:Number-of-trajectories}}
\end{table}

For a better comparability between spectra, we always substract the
uncoupled bath ground state energy in our plots, i. e.,
\begin{eqnarray}
E_{\text{plot}} & = & E-\sum\limits _{i=0}^{F_{\text{bath}}}\omega_{i}/2.
\end{eqnarray}

\begin{figure}
\begin{centering}
\includegraphics[scale=0.4]{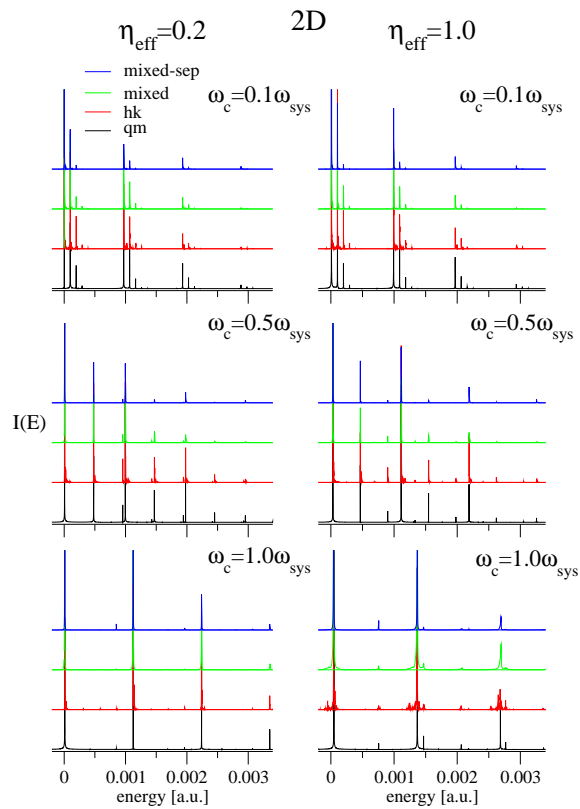}
\par\end{centering}

\caption{\label{fig:2D}Two dimensional power spectra calculation at different
level of semiclassical accuracy for different cutting frequency and
bath friction values.}
\end{figure}
Figure \ref{fig:2D} shows the power spectra comparison for the two-dimensional
simulations. For each plot, exact quantum wavepacket propagation is
compared with the HK SC-IVR calculations at different levels of approximation.
The ``hk'' label is for the Herman-Kluk SC-IVR propagator, the ``mixed''
is for Eq. (\ref{eq:sep_approx-5}) and the `` mixed-sep'' is for
Eq. (\ref{eq:sep_B}). To better appreciate the different levels of
approximation, the results should be read in a hierarchical order
by comparing the ``mixed-sep'' with the ``mixed'', the ``mixed''
with the ``hk'' and the ``hk'' with the exact ``qm''. It is
quite surprising that independently of the coupling and of the system
frequency, the separable approximation of Eq. (\ref{eq:sep_B}), labeled
``mixed-sep'', faithfully reproduces the spectral profile of the
original approximation of Eq. (\ref{eq:sep_approx-5}), the ``mixed''
one. Peaks locations are also well reproduced by the ``mixed'' approximation
with respect to the original ``hk'' one. However, the peak intensities
are not well reproduced in the different approximations. Thus, the
approximations described in Sections \ref{sec:Mixed} and \ref{sec:Sep-Mixed}
are suitable for locating the vibrational eigenvalues but it introduces
some inaccuracy in the spectral intensities. Finally, the difference
between the ``hk'' spectrum and the exact one are only for peak
intensities and the different time propagation. In fact, the ``hk''
is employing long simulation times to exploit the time-averaging filter,
while the quantum wave packet simulations are stopped at some computationally
feasible time. All three cases show a blueshift of the system frequency
(very clear in the middle panel) and a redshift of the bath frequency.
For the stronger coupling, the blueshift tendency of the system is
enhanced.

\begin{figure}
\begin{centering}
\includegraphics[scale=0.4]{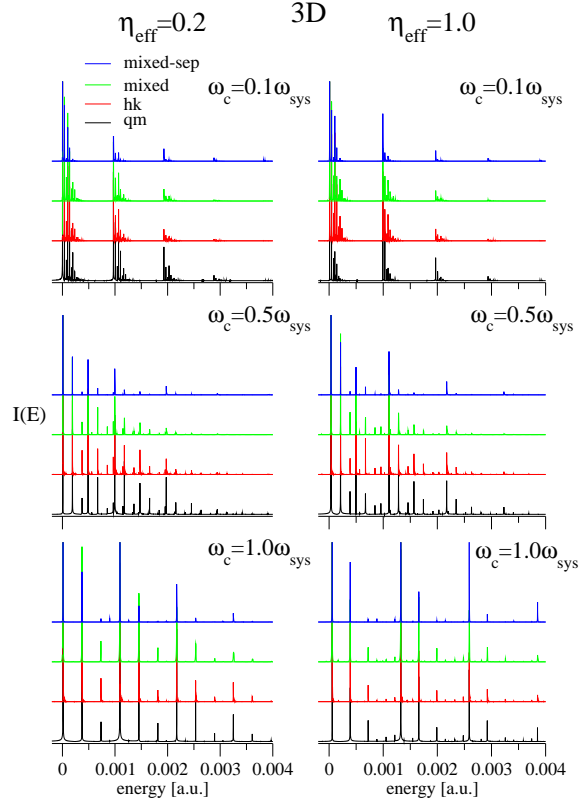}
\par\end{centering}

\caption{\label{fig:3D}As in Fig.(\ref{fig:2D}), but for three dimensional
calculations.}
\end{figure}

We now turn to the three-dimensional calculations shown in Fig. \ref{fig:3D}.
Here, the spectroscopic features are much more complex and the simulation
is quite more challenging than the previous one. Nevertheless, the
``mixed'' approximation is able to reproduce the ``hk'' results
quite faithfully, since the peak positions are correct both in the
resonant and off-resonance case. The most severe ``mixed-sep'' approximation
is also reproducing both the ``hk'' and ``mixed'' results, even
if it introduces some inaccuracyin the peak intensity and a few highly
excited overtones are missing. Also in three dimensions, for the resonant
case $\omega_{\text{c}}=\omega_{\text{s}}$, the resonant bath frequencies
are strongly red shifted ,while the blue shift of the system is further
enhanced.

Considering the drastic computational effort reduction introduced
by the Gaussian integration in Eq. (\ref{eq:sep_approx-5}), where
a phase space integral is approximated by a single phase space trajectory,
we think that the results in Fig.s (\ref{fig:2D}) and (\ref{fig:3D})
are quite satisfying. Even when the coupling between the system and
the bath is resonant, the peak position is still quite accurate. The
main drawback of the approximations ``mixed'' and ``mixed-sep''
is represented by the loss of intensity for some peaks. However, for
complex systems this limitation could help for a better peak interpretation
because the ``mixed-sep'' approximation mainly suppresses bath excitations
in the spectrum, as we show in Appendix B.

On one hand, the ``mixed'' approximation is more computational expensive
per trajectory than the time averaged HK calculation because of the
double time-integration. On the other hand, the number of trajectories
needed for converging the phase space integral is reduced with respect
to the full dimensional HK integration, because the integration is
limited to the system's phase space. Table 1 shows that the ``hk''
calculation might still be faster than the full ``mixed'' one in
spite of the reduction in the number of trajectories needed for convergence
because of the unfavorable scaling of the ``mixed'' approximation
with the number of time steps and because the calculation of the classical
trajectories does not take much time in this example. For a realistic
problem where the potential is not given analytically and the trajectory
is simulated on-the-fly \cite{Ceotto_1traj,Ceotto_acceleratedSCIVR,Ceotto_cursofdimensionality_11,Ceotto_MCSCIVR,Ceotto_NH3},
the trajectory calculation will take much longer and make the ``mixed''
approach the more efficient one. The ``mixed-sep'' approximation,
on the other hand, combines the single time integration of ``hk''
with the reduced number of trajectories, which results in an impressive
speedup of the computations, as shown in Table 1. The inaccuracy that
comes with this additional approximation is only in peak intensity,
but not peak position, as discussed before.

\section{Conclusions\label{sec:Conclusions}}

We have developed a new semiclassical method for the calculation of
vibrational spectra of molecular systems that is based on the combination
of the time-averaging idea with the semiclassical hybrid methodology.
After partitioning the phase space variables into system and bath
ones, we could apply a lower accuracy semiclassical propagation scheme
based on a thawed Gaussian approximation to the bath degrees of freedom
only, while preserving the full HK semiclassical propagator accuracy
for the system time evolution. The resulting expression of the power
spectrum intensity of Eq. (\ref{eq:sep_approx-5}) was then approximated
to further reduce the computational effort. In this way, the separabale
approximation, that is at the heart of the time-averaging method,
was implemented and lead to the final working formula, given in Eq.
(\ref{eq:sep_B}). In the harmonic case, this additional approximation
has been shown to give identical peak positions in the spectrum, as
can be seen by comparing the results in Appendix A and Appendix B.
Numerical examples have been shown for a Morse oscillator coupled
to one or two harmonic oscillators with different frequencies and
coupling strengths. Even in the case of only one additional harmonic
degree of freedom, compared to full Herman-Kluk time-averaging results,
the numerical effort was shown to be reduced by more than an order
of magnitude. Considering different system-bath couplings and including
the resonant scenario, we found peak positions to be always in very
good agreement with full quantum calculations, that are still feasible
in the cases considered.In the future, we are planning to apply this
semiclassical method to more realistic systems, where the harmonic
bath is replaced by a realistic solvent, such as rare gas matrices
\cite{Buchholz_2011,BiharyApkarian_2001} or even water molecules.

\section*{Acknowledgement}

Michele Ceotto acknowledges support from the European Research Council
(ERC) under the European Union\textquoteright{}s Horizon 2020 research
and innovation programme (grant agreement No {[}647107{]} \textendash{}
SEMICOMPLEX \textendash{} ERC-2014-CoG). M.C. acknowledges also the
CINECA and the Regione Lombardia award under the LISA initiative (grant
SURGREEN) for the availability of high performance computing resources
and the Chemistry Department of the University of Milan for funding
through the Development Plan of Athenaeum grant \textendash{} line
B1 (UNIAGI 17777). FG gratefully acknowledges financial support from
the Deutsche Forschungsgemeinschaft through Grant No. GR 1210/4-2.

\section*{Appendix A: The Harmonic Spectrum using a Thawed Gaussian Wavepacket\label{sec:Appendix-A:TWD}}

In this appendix, we show that the thawed Gaussian approximation (TGA)
as an exact solution of the time-dependent Schrödinger equation for
the harmonic oscillator leads to the exact harmonic spectrum. The
wavepacket in the TGA \cite{Heller_thawedgaussian} is written as
the coherent state
\begin{equation}
\psi\left(x,t\right)=\left(\frac{\gamma_{0}}{\pi}\right)^{1/4}\exp\left\{ -\frac{\gamma_{t}}{2}\left(x-q_{t}\right)^{2}+\frac{\text{i}}{\hbar}p_{t}\left(x-q_{t}\right)+\frac{\text{i}}{\hbar}\delta_{t}\right\} ,\label{eq:wavepacket_TWGD}
\end{equation}
where the parameters $\gamma_{t}$ and $\delta_{t}$ evolve with time
according to the differential equations
\begin{eqnarray}
-\text{i}\hbar\dot{\gamma}_{t} & = & -\frac{\hbar^{2}}{m}\gamma_{t}^{2}+\frac{\text{d}{}^{2}}{\text{d}q_{t}^{2}}V\left(q_{t},t\right)\label{eq:gamma_TWGD}\\
\dot{\delta}_{t} & = & \frac{p_{t}^{2}}{2m}-V\left(q_{t},t\right)-\frac{\hbar^{2}}{2m}\gamma_{t}.\label{eq:delta_TWGD}
\end{eqnarray}
For the harmonic oscillator motion where the potential is $V\left(q_{t}\right)=m\omega^{2}q{}_{t}^{2}/2$,
(\ref{eq:gamma_TWGD}) and (\ref{eq:delta_TWGD}) become
\begin{eqnarray}
-\text{i}\hbar\dot{\gamma}_{t} & = & -\frac{\hbar^{2}}{m}\gamma_{t}^{2}+m\omega^{2}\label{eq:gamma2_TWDG}\\
\delta_{t} & = & S_{t}\left(p_{0},q_{0}\right)-\frac{\hbar\omega}{2}t,\label{eq:delta2_TWDG}
\end{eqnarray}
and $\gamma\left(t\right)$ is constant for the arbitrary and convenient
choice of $\gamma\left(0\right)=m\omega/\hbar$. In this case, (\ref{eq:wavepacket_TWGD})
becomes
\begin{equation}
\psi\left(x,t\right)=\left(\frac{\gamma_{0}}{\pi}\right)^{1/4}\exp\left\{ -\frac{\gamma_{0}}{2}\left(x-q_{t}\right)^{2}+\frac{\text{i}}{\hbar}p_{t}\left(x-q_{t}\right)+\frac{\text{i}}{\hbar}S_{t}-\frac{\text{i}}{\hbar}\left(\frac{\hbar\omega}{2}\right)t\right\} .\label{eq:wavepacket_t_TWGD}
\end{equation}
The power spectrum is thus given by
\begin{eqnarray}
I(E) & = & \frac{1}{2\pi\hbar}\int_{-\infty}^{+\infty}\text{d}t\text{e}^{\text{i}Et/\hbar}\left\langle \psi\left(0\right)|\psi\left(t\right)\right\rangle \nonumber \\
 & = & \frac{1}{2\pi\hbar}\int_{-\infty}^{+\infty}\text{d}t\text{e}^{\text{i}Et/\hbar}\int_{-\infty}^{+\infty}\psi^{*}\left(x,0\right)\psi\left(x,t\right)\text{d}x\nonumber \\
 & = & \frac{1}{2\pi\hbar}\int_{-\infty}^{+\infty}\text{d}t\exp\left\{ \frac{\text{i}}{\hbar}\left(E-\frac{\hbar\omega}{2}\right)t\right\} \label{eq:AppendixA_PowerSpectrum_I(E)}\\
 &  & \times\exp\left\{ -\frac{\gamma_{0}}{4}\left(q_{t}-q_{0}\right)^{2}-\frac{1}{4\hbar^{2}\gamma_{0}}\left(p_{t}-p_{0}\right)^{2}+\frac{\text{i}}{2\hbar}\left(q_{0}p_{t}-q_{t}p_{0}\right)\right\} .\nonumber
\end{eqnarray}
After inserting the harmonic oscillator solutions
\begin{eqnarray}
p_{t} & = & p_{0}\cos\omega t-m\omega q_{0}\sin\omega t\label{eq:AppendixA_ptHO}\\
q_{t} & = & q_{0}\cos\omega t+\frac{p_{0}}{m\omega}\sin\omega t,\label{eq:AppendixA_qtHO}
\end{eqnarray}
choosing $\gamma_{0}=m\omega/\hbar$ and doing a power series expansion
of the second part of the exponential, the time integration in (\ref{eq:AppendixA_PowerSpectrum_I(E)})
can be done analytically to yield
\begin{eqnarray}
I(E) & = & \frac{1}{2\pi\hbar}\int_{-\infty}^{\infty}\text{d}t\ \exp\left\{ \frac{\text{i}}{\hbar}\left[\left(E-\frac{\hbar\omega}{2}\right)t\right]+\frac{m\omega}{2\hbar}\left(\frac{p_{0}^{2}}{m^{2}\omega^{2}}+q_{0}^{2}\right)\left(\text{e}^{-\text{i}\omega t}-1\right)\right\} \label{eq:AppendixA_PowerSpectrum_I(E)_1}\\
 & = & \exp\left\{ -\frac{m\omega}{2\hbar}q_{0}^{2}-\frac{1}{2m\omega\hbar}p_{0}^{2}\right\} \label{eq:AppendixA_PowerSpectrum_I(E)_2}\\
 &  & \times\sum_{k=0}^{+\infty}\frac{1}{2^{k}k\text{!}}\left(\frac{m\omega}{\hbar}q_{0}^{2}+\frac{p_{0}^{2}}{m\omega\hbar}\right)^{k}\frac{1}{2\pi\hbar}\int_{-\infty}^{\infty}\text{d}t\ \exp\left\{ \frac{\text{i}}{\hbar}\left[\left(E-\frac{\hbar\omega}{2}-\hbar\omega k\right)t\right]\right\} \nonumber \\
 & = & \exp\left\{ -\frac{m\omega q_{0}^{2}}{2\hbar}-\frac{p_{0}^{2}}{2m\omega\hbar}\right\} \sum_{k=0}^{+\infty}\frac{1}{2^{k}k\text{!}}\left(\frac{m\omega q_{0}^{2}}{\hbar}+\frac{p_{0}^{2}}{m\omega\hbar}\right)^{k}\delta\left(E-\hbar\omega\left(k+\frac{1}{2}\right)\right),\label{eq:AppendixA_PowerSpectrum_I(E)_3}
\end{eqnarray}
which is the full harmonic spectrum, i.e. all vibrational levels are
exactly reproduced.

\section*{Appendix B: The Harmonic Spectrum From the Hybrid Expressions\label{sec:Appendix-B:TWD-separable}}

We demonstrate in this appendix how the approximations of the TG part
leading from the original mixed semiclassical expression (\ref{eq:sep_approx-5})
to the simplified formula (\ref{eq:sep_B}) affect the harmonic oscillator
spectrum. In order to do so, we first go through the basic steps of
calculating the spectrum (\ref{eq:AppendixA_PowerSpectrum_I(E)_3})
from Eq. (\ref{eq:sep_approx-5}). To keep it simple, we show this
for one harmonic oscillator DOF that is treated with the TGA. The
expression for the spectrum then emerges as a limiting case of Eq.
(\ref{eq:sep_approx-5}) and reads
\begin{eqnarray}
I\left(E\right) & = & \frac{1}{2\hbar}\frac{\text{Re}}{\pi\hbar T}\int_{0}^{T}\text{d}t_{1}\int_{t_{1}}^{T}\text{d}t_{2}\ \text{e}^{\text{i}\left[E\left(t_{1}-t_{2}\right)+\phi_{t_{1}}\left(p_{0},q_{0}\right)-\phi_{t_{2}}\left(p_{0},q_{0}\right)\right]/\hbar}\nonumber \\
 & \times & \sqrt{\frac{1}{\det\left(\mathbf{A}_{t_{1}}+\mathbf{A}_{t_{2}}^{*}\right)}}\mbox{exp}\left\{ \frac{1}{4}\left(\mathbf{b}_{t_{1}}+\mathbf{b}_{t_{2}}^{*}\right)^{\text{T}}\left(\mathbf{A}_{t_{1}}+\mathbf{A}_{t_{2}}^{*}\right)^{-1}\left(\mathbf{b}_{t_{1}}+\mathbf{b}_{t_{2}}^{*}\right)+c_{t_{1}}+c_{t_{2}}^{*}\right\} .\label{eq:sep_approx-5-1D}
\end{eqnarray}
Using $p_{t}$ and $q_{t}$ from (\ref{eq:AppendixA_ptHO}) and (\ref{eq:AppendixA_qtHO})
and calculating the respective derivatives, the matrix $\mathbf{A}_{t}$
from equation (\ref{eq:A_matrix}) becomes a constant in time
\begin{eqnarray}
\mathbf{A} & = & \left(\begin{array}{cc}
1/(4m\omega\hbar) & \text{i}/(4\hbar)\\
\text{i}/(4\hbar) & m\omega/(4\hbar)
\end{array}\right)
\end{eqnarray}
and the determinant from the prefactor is
\begin{eqnarray}
\det\left(\mathbf{A}_{t_{1}}+\mathbf{A}_{t_{2}}^{*}\right) & = & \frac{1}{4\hbar^{2}}.
\end{eqnarray}
The vector $\mathbf{b}_{t}$, defined in (\ref{eq:B_definition}),
(\ref{eq:b1}), and (\ref{eq:b2}), has components
\begin{eqnarray}
b_{1,t} & = & \frac{1}{2\hbar}\left(\text{e}^{-\text{i}\omega t}-1\right)\left(\frac{p_{0}}{m\omega}+\text{i}q_{0}\right)\\
b_{2,t} & = & -\text{i}\frac{m\omega}{2\hbar}\left(\text{e}^{-\text{i}\omega t}-1\right)\left(\frac{p_{0}}{m\omega}+\text{i}q_{0}\right)-\frac{\text{i}}{\hbar}p_{0},
\end{eqnarray}
making the first term of the exponent in (\ref{eq:sep_approx-5-1D})
\begin{eqnarray}
\frac{1}{4}\left(\mathbf{b}_{t_{1}}+\mathbf{b}_{t_{2}}^{*}\right)^{\text{T}}\left(\mathbf{A}_{t_{1}}+\mathbf{A}_{t_{2}}^{*}\right)^{-1}\left(\mathbf{b}_{t_{1}}+\mathbf{b}_{t_{2}}^{*}\right) & = & \frac{m\omega}{2\hbar}\left(\frac{p_{0}^{2}}{m^{2}\omega^{2}}+q_{0}^{2}\right)\label{eq:AppendixB_exponent_bAb}\\
 &  & \times\left(\text{e}^{\text{i}\omega\left(t_{2}-t_{1}\right)}-\text{e}^{-\text{i}\omega t_{1}}-\text{e}^{\text{i}\omega t_{2}}+1\right).\nonumber
\end{eqnarray}
With the harmonic oscillator action
\begin{eqnarray}
S_{t} & = & \left(\frac{p_{0}^{2}}{2m\omega}-\frac{1}{2}m\omega q_{0}^{2}\right)\cos\omega t\sin\omega t-p_{0}q_{0}\sin^{2}\omega t,
\end{eqnarray}
the scalar $c_{t}$ from equation (\ref{eq:c_t}) is
\begin{eqnarray}
c_{t} & = & \frac{m\omega}{2\hbar}\left(\frac{p_{0}^{2}}{m^{2}\omega^{2}}+q_{0}^{2}\right)\left(\text{e}^{-\text{i}\omega t}-1\right).\label{eq:AppendixB_exponent_ct}
\end{eqnarray}
Adding $c_{t_{1}}+c_{t_{2}}^{*}$ to (\ref{eq:AppendixB_exponent_bAb}),
the total exponent in (\ref{eq:sep_approx-5-1D}) is found to be
\begin{eqnarray}
\frac{1}{4}\left(\mathbf{b}_{t_{1}}+\mathbf{b}_{t_{2}}^{*}\right)^{\text{T}}\left(\mathbf{A}_{t_{1}}+\mathbf{A}_{t_{2}}^{*}\right)^{-1}\left(\mathbf{b}_{t_{1}}+\mathbf{b}_{t_{2}}^{*}\right)+c_{t_{1}}+c_{t_{2}}^{*} & = & \frac{m\omega}{2\hbar}\left(\frac{p_{0}^{2}}{m^{2}\omega^{2}}+q_{0}^{2}\right)\\
 &  & \times\left(\text{e}^{\text{i}\omega\left(t_{2}-t_{1}\right)}-1\right).\nonumber
\end{eqnarray}
Taking into account the phase of the prefactor $\phi(t)=-\hbar\omega t/2$
for the harmonic oscillator, the total expression for the spectrum
takes the form
\begin{eqnarray}
I\left(E\right) & = & \frac{\text{Re}}{\pi\hbar T}\int_{0}^{T}\text{d}t_{1}\int_{t_{1}}^{T}\text{d}t_{2}\ \exp\left\{ \frac{\text{i}}{\hbar}\left[E\left(t_{1}-t_{2}\right)-\frac{\hbar\omega}{2}\left(t_{1}-t_{2}\right)\right]\right\} \\
 &  & \times\exp\left\{ \frac{m\omega}{2\hbar}\left(\frac{p_{0}^{2}}{m^{2}\omega^{2}}+q_{0}^{2}\right)\left(\text{e}^{-\text{i}\omega\left(t_{1}-t_{2}\right)}-1\right)\right\} .
\end{eqnarray}
Changing variables to $\tau\equiv t_{2}-t_{1}$ yields after integration
over $\tau_{2}=t_{1}$
\begin{eqnarray}
I\left(E\right) & = & \frac{\text{Re}}{\pi\hbar T}\int_{0}^{T}\text{d}\tau\ \exp\left\{ -\frac{\text{i}}{\hbar}\left[\left(E-\frac{\hbar\omega}{2}\right)\tau\right]+\frac{m\omega}{2\hbar}\left(\frac{p_{0}^{2}}{m^{2}\omega^{2}}+q_{0}^{2}\right)\left(\text{e}^{\text{i}\omega\tau}-1\right)\right\} ;\label{eq:AppendixB_I(E)_1DHO_full}
\end{eqnarray}
a series expansion of the second part of the exponent and another
integration in the limit $T\rightarrow\infty$ reproduces (\ref{eq:AppendixA_PowerSpectrum_I(E)_1}).

Second, we consider the simplified hybrid approximation (\ref{eq:sep_B}),
which has the 1D TGA form
\begin{eqnarray}
I\left(E\right) & = & \frac{1}{2\hbar}\frac{1}{2\pi\hbar T}\left|\int_{0}^{T}\text{d}t\:\text{e}^{\text{i}\left[Et+\phi_{t}\left(p_{0},q_{0}\right)\right]/\hbar}\right.\label{eq:Appendix_B_I(E)_1DHO_approx_1}\\
 &  & \times\frac{1}{\left[\det\left(\mathbf{A}_{t}+\mathbf{A}_{t}^{*}\right)\right]^{1/4}}\left.\exp\left\{ \frac{1}{4}\mathbf{b}_{\text{m},t}^{\text{T}}\left(\mathbf{A}_{t}+\mathbf{A}_{t}^{*}\right)^{-1}\mathbf{b}_{\text{m},t}+c_{t}\right\} \right|^{2}.\nonumber
\end{eqnarray}
The prefactor phase $\phi_{t}$ and scalar term in the exponent $c_{t}$
obviously stay the same as in the full expression. Due to its time
independence for the harmonic oscillator, the terms containing $\mathbf{A}_{t}$
do not change either. This makes the approximation of the determinant
(\ref{eq:pre-exp_approx}) an exact identity. Comparing (\ref{eq:AppendixB_exponent_ct}),
(\ref{eq:AppendixB_I(E)_1DHO_full}) and (\ref{eq:Appendix_B_I(E)_1DHO_approx_1}),
we see that $c_{t}$ is the only contribution we need for the exponent.
Consequently, the modified vector $\mathbf{b}_{\text{m},t}$, where
the second component no longer contains the constant imaginary part,
\begin{eqnarray}
b_{\text{m},2,t}^{\text{T}} & = & -\text{i}\frac{m\omega}{2\hbar}\left(\text{e}^{-\text{i}\omega t}-1\right)\left(\frac{p_{0}}{m\omega}+\text{i}q_{0}\right),
\end{eqnarray}
is designed such that the contributions from the two components of
$\mathbf{b}_{\text{m},t}$ in the exponent cancel each other,
\begin{eqnarray}
\frac{1}{4}\mathbf{b}_{\text{m},t}^{\text{T}}\left(\mathbf{A}_{t}+\mathbf{A}_{t}^{*}\right)^{-1}\mathbf{b}_{\text{m},t} & = & 0,
\end{eqnarray}
and the power spectrum resulting from these approximations
\begin{eqnarray}
I\left(E\right) & = & \frac{1}{2\pi\hbar T}\left|\int_{0}^{T}\text{d}t\:\exp\left\{ \frac{\text{i}}{\hbar}\left(Et-\frac{\hbar\omega}{2}t\right)+\frac{m\omega}{2\hbar}\left(\frac{p_{0}^{2}}{m^{2}\omega^{2}}+q_{0}^{2}\right)\left(\text{e}^{-\text{i}\omega t}-1\right)\right\} \right|^{2}
\end{eqnarray}
has a time integrand that is identical to the one in the full expression
from Eq. (\ref{eq:AppendixB_I(E)_1DHO_full}). After a series expansion
of the exponential as before, unfolding the modulos into the double
integral $2\text{Re}\int_{0}^{T}dt_{1}\int_{t_{1}}^{T}dt_{2}$ and
changing variables as suggested above, the result after time integration
is
\begin{eqnarray}
I\left(E\right) & = & \exp\left\{ -\frac{m\omega q_{0}^{2}}{\hbar}-\frac{p_{0}^{2}}{m\omega\hbar}\right\} \sum_{k=0}^{+\infty}\frac{1}{(k!)^{2}}\left(\frac{m\omega q_{0}^{2}}{2\hbar}+\frac{p_{0}^{2}}{2m\omega\hbar}\right)^{2k}\delta\left(E-\hbar\omega\left(k+\frac{1}{2}\right)\right).
\end{eqnarray}
All harmonic oscillator peaks are placed at the right positions. Only
the relative peak weight is the squared value compared to the correct
result, thus damping peaks from higher (bath) excitations.


\begin{thebibliography}{References}
\bibitem{Miller_avd_74}(a) W. H. Miller, J. Chem. Phys. \textbf{53},
3578 (1970); (b)\emph{ ibidem} \textbf{53}, 1949 (1970); (c) W. H.
Miller, J. Phys. Chem. A \textbf{105}, 2942 (2001).

\bibitem{Heller_IVR}(a) E. J. Heller, J. Chem. Phys. \textbf{62},
1544 (1975); (b) \emph{ibidem} \textbf{75}, 2923 (1981); (c) ibidem
\textbf{\noun{94}}, 2723 (1991).

\bibitem{Miller_PNAS}W. H. Miller, Proc. Natl. Acad. Sci. U.S.A.
\textbf{102}, 6660 (2005).

\bibitem{Kay_review}K. G. Kay, Annu. Rev. Phys. Chem. \textbf{56},
255 (2005).

\bibitem{Pollak_perturbationSeries}S. S. Zhang and E. Pollak, J.
Chem. Phys. \textbf{121}, 3384 (2004).

\bibitem{Herman_Kluk_2} M. F. Herman, J. Chem. Phys. 85, 2069 (1986);
E. Kluk, M. F. Herman and H. L. Davis, J. Chem. Phys. 84, 326 (1986)

\bibitem{Pollak_prefactorfree_05} S. Zhang and E. Pollak, J. Chem.
Theory Comput. \textbf{1}, 345 (2005).

\bibitem{Apkarian}M. Ovchinnikov and V. A. Apkarian, J. Chem. Phys.
105 (23), 10312 (1996); \emph{ibidem} 106 (13), 5775 (1997); \emph{ibidem}
108 (6), 2277 (1998).

\bibitem{MIller_Sun_mixedSCclassical}X. Sun and W. H. Miller, J.
Chem. Phys. \textbf{130}106 (3), 916 (1997).

\bibitem{Miller336_GeneralizedFBSCIVR_01}M. Thoss, H. Wang and W.
H. Miller, J. Chem. Phys. \textbf{114}, 9220 (2001)

\bibitem{Nandini_mixed_quantum_classical}S. V. Antipov, Z. Ye, and
N. Ananth, J. Chem. Phys. \textbf{142}, 184102 (2015)

\bibitem{Grossmann_hybrid_06} F. Grossmann, J. Chem. Phys. \textbf{125},
014111 (2006).

\bibitem{Heller_thawedgaussian} E. J. Heller, J. Chem. Phys. \textbf{62},
1544 (1975).

\bibitem{Grossmann_thwg}F. Grossmann, Comments At. Mol. Phys. \textbf{34},
141 (1999).

\bibitem{Deshpande_Ezra_2006-1}S. A. Deshpande, G. S. Ezra, J. Phys.
A 39, 5067 (2006).

\bibitem{Jiri_oligotiophenes_14} M. Wehrle, M. Sulc, J. Vanicek,
J. Chem. Phys. \textbf{140}, 244114 (2014)

\bibitem{Jiri_ammonia_2015}M. Wehrle, S. Oberli, J. Vanicek, J. Phys.
Chem A \textbf{119}, 5685 (2015)

\bibitem{Jiri_Miroslav_MolPHys_2012}S. Miroslav and J. Vanicek, Molecular
Physics \textbf{110}, 945 (2012)

\bibitem{Pollak_Conte_TGWD}R. Conte and E. Pollak, Phys. Rev. E \textbf{81},
036704 (2010).

\bibitem{Kay-TA}Y. Elran and K. G. Kay, J. Chem. Phys.\emph{ }\textbf{110},
3653 (1999); \emph{ibidem} \textbf{110}, 8912 (1999).

\bibitem{Alex_Mik}A. L. Kaledin and W. H. Miller, J. Chem. Phys.
\textbf{118}, 7174 (2003); A. L. Kaledin and W. H. Miller, J. Chem.
Phys. \textbf{119}, 3078 (2003).

\bibitem{Heller_review_autocorrel}E. J. Heller, Acc. Chem. Res. \textbf{14},
368 (1981)

\bibitem{Heller_frozengaussian} E. J. Heller, J. Chem. Phys. 75,
2923 (1981).

\bibitem{Kay_94} (a) K. G. Kay, J. Chem. Phys. \textbf{100}, 4377
(1994); (b) \emph{ibidem} \textbf{100}, 4432 (1994).; (c) K. G. Kay,
J. Chem. Phys. \textbf{101}, 2250 (1994).

\bibitem{Grossmann-Xavier}F. Grossmann and A. L. Xavier, Phys. Lett.
243, 243 (1998).

\bibitem{Manolopoulos} (a) A. R. Walton, D. E. Manolopoulos, Mol.
Phys. \textbf{87}, 961 (1996); (b) A. R: Walton, D. E. Manolopoulos,
Chem. Phys. Lett\emph{.} \textbf{244}, 448 (1995); (c) M. L. Brewer,
J. S. Hulme, D. E. Manolopoulos, J. Chem. Phys\emph{.} \textbf{106},
4832 (1997).

\bibitem{Coker}S. Bonella, D. Montemayor , and D. F. Coker, Proc.
Natl. Am. Soc\emph{.} \textbf{102}, 6715 (2005).

\bibitem{Takatsuka_eigenstates_05} (a) H. Ushiyama and K. Takatsuka,
J. Chem. Phys. \textbf{122}, 224112 (2005); (b) S. Takahashi and K.
Takatsuka, J. Chem. Phys. \textbf{127}, 084112 (2007).

\bibitem{Roy}B. B. Issack and P. N. Roy, J. Chem. Phys. \textbf{127},
054105 (2007).

\bibitem{Pollak_autocorrelation} (a) E. Pollak and E. Martin-Fierro,
J. Chem. Phys. \textbf{126}, 164107 (2007); (b) E. Martin-Fierro and
E. Pollak, J. Chem. Phys. \textbf{125}, 164104 (2006).

\bibitem{Miller_vari} (a) H. Wang, X. Sun, and W. H. Miller, J. Chem.
Phys. \textbf{108}, 9726 (1988); (b) T. Yamamoto, H. Wang, and W.
H. Miller, J. Chem. Phys. \textbf{116}, 7335 (2002); (c) T. Yamamoto,
W. H. Miller, J. Chem. Phys. \textbf{118}, 2135 (2003).

\bibitem{Miller_Tao_timedependentsampling_11} G. Tao and W. H. Miller,
J. Chem. Phys. \textbf{135}, 024104 (2011); \emph{ibidem} \textbf{137},
124105 (2012); G. Tao and W. H. Miller, J. Chem. Phys. \textbf{137},
124105 (2012)

\bibitem{Roy_AbinitioSCIVR}S. Y. Y. Wong, D. M. Benoit, M. Lewerenz,
A. Brown, and P.-N. Roy, J. Chem. Phys. \textbf{134}, 094110 (2011)

\bibitem{Maitra_SCmaps_00_filter}N. T. Maitra, J. Chem. Phys. \textbf{112},
531 (2000).

\bibitem{Liu_IJQC_review_15}J. Liu, Int. J. of Quantum Chemistry,
\textbf{115} (11), 657 (2015)

\bibitem{Tao_nonadaJPCA_13}G. Tao, J. Phys. Chem. A \textbf{117},
5821\textendash{}5825 (2013)

\bibitem{Burant_Batista}J.C. Burant and V.S. Batista, J. Chem. Phys.
\textbf{116}, 2748\textendash{}2756 (2002)

\bibitem{Ceotto_MCSCIVR} M. Ceotto, S. Atahan, G. F. Tantardini,
and A. Aspuru-Guzik,\emph{ }J. Chem. Phys. \textbf{130}, 234113, (2009).

\bibitem{Ceotto_1traj}M. Ceotto, S. Atahan, S. Shim, G. F. Tantardini,
and A. Aspuru-Guzik, Phys. Chem. Chem. Phys\emph{.} \textbf{11}, 3861
(2009).

\bibitem{Ceotto_eigenfunctions}M. Ceotto, S. Valleau, G. F. Tantardini,
and A. Aspuru-Guzik, J. Chem Phys. \textbf{134}, 234103 (2011).

\bibitem{Ceotto_cursofdimensionality_11}M. Ceotto, G. F. Tantardini,
and A. Aspuru-Guzik, J. Chem. Phys. \textbf{135}, 214108 (2011).

\bibitem{Ceotto_david}M. Ceotto, D. dell'Angelo, and G. F. Tantardini,
J. Chem. Phys. \textbf{133}, 054701 (2010).

\bibitem{Ceotto_acceleratedSCIVR}M. Ceotto, Y. Zhuang, and W. L.
Hase, J. Chem. Phys. \textbf{138}, 054116 (2013).

\bibitem{Ceotto_NH3}R. Conte, A. Aspuru-Guzik, and M. Ceotto, \emph{J.
Phys. Chem. Lett.} \textbf{4}, 3407 (2013)

\bibitem{Ceotto_Zhang_JCTC} Y. Zhuang, M. R. Siebert, W. L. Hase,
K. G. Kay, and M. Ceotto, J. Chem. Theory and Comput \textbf{9}, 54
(2013)

\bibitem{Ceotto_GPU} D. Tamascelli, F. S. Dambrosio, R. Conte, and
M. Ceotto, J. Chem. Phys. \textbf{140}, 174109 (2014)

\bibitem{CaldeiraLeggett_1983} A. O. Cladeira and A. J. Leggett,
Physica A \textbf{121}, 587 (1983)

\bibitem{Goletz_2009}C. Goletz, and F. Grossmann, J. Chem. Phys.
\textbf{130}, 244107 (2009)

\bibitem{SchmidtLorenz2011} B. Schmidt and U. Lorenz, WavePacket
4.7.3, available via http://sourceforge.net/projects/wavepacket/ (2011)

\bibitem{Buchholz_2011} M. Buchholz, C. Goletz, F. Grossmann, B.
Schmidt, J. Heyda, and P. Jungwirth, J. Phys. Chem. A, \textbf{116},
11199 (2012)

\bibitem{BiharyApkarian_2001} Z. Bihary, R. B. Gerber, and V. A.
Apkarian, J. Chem. Phys. \textbf{115}, 2695 (2001)\end{thebibliography}
\end{document}